**Conceptualizing Smart City Applications: Requirements, Architecture, Security Issues and Emerging Trends**

**A. K. M. Bahalul Haque**

**Bharat Bhushan**

**Gaurav Dhiman**

## Abstract

The emergence of smart cities and sustainable development has become a globally accepted form of urbanization. The epitome of smart city development has become possible due to the latest innovative integration of information and communication technology. Citizens of smart cities can enjoy the benefits of a smart living environment, ubiquitous connectivity, seamless access to services, intelligent decision making through smart governance, and optimized resource management. The widespread acceptance of smart cities has raised data security issues, authentication, unauthorized access, device-level vulnerability, and sustainability. This paper focuses on the wholistic overview and conceptual development of smart city. Initially, the work discusses the smart city idea and fundamentals explored in various pieces of literature. Further various smart city applications, including notable implementations, are put forth to understand the quality of living standards. Finally, the paper depicts a solid understanding of different security and privacy issues, including some crucial future research directions.

### 1. Introduction

A smart city [1] is the idea of creating a sustainable living environment along with state-of-the-art technology (ICT) integration. The smart city citizens will have seamless and ubiquitous access to information that allows them to control their lives using collective cyber intelligence. The city will provide ICT-based innovative and sustainable solutions by utilizing natural and economic resources [2]. The smart city concept includes a higher quality of life(QoL) for citizens that offer smart healthcare and institutes, decentralized economy and government, intelligent vehicles, etc.

Over the past couple of decades, the population has increased, and more than 50% live in urban areas. The estimation suggests the number shall rise to 70% within the next three decades [3]. Rapid population increase poses a few scenarios. The increasing number of the population needs additional basic human needs. These demands must be fulfilled to establish a stable society and economy. The resources are limited throughout the world. With these limited resources, production and distribution have to be at an optimum level. There are some renewable resources, for example, renewable energy. Though the energy source is renewable, smart management needs to be adopted. The same goes for the supply chain, healthcare, real estate, transportation, etc. A smart city idea offers a suitable smart solution for all those applications mentioned above. The idea is to provide sufficient food [4], energy [5], state of the art transportation [6], efficient supply chain management system [7], efficient infrastructure [8], smart data management [9], etc.

If the basic human need, including the other amenities, can be provided to the citizens, this will lead to a productive human resource, an effective working environment, good governance, etc. [10]. A smart city is a self-containing city that focuses on people's quality of life above everything else. The goal of a smart city is multifaceted. Smart citizens [11] should have digital identities, enabling them to share data that can be processed to improve the quality of services. Smart city demands sustainability that is the composition of a greener lifestyle and comfortable social life. Reducing environmental pollution and proper waste management preserves the smart city ecosystem. The governance and economy should be decentralized and completely transparent. Besides, different types of security attacks can compromise smart city applications and infrastructure. Citizens of a smart city can lose data. Various organizations like banks, insurance companies, hospitals, etc. can potentially be victims of data loss and face a massive crisis. Everything in a smart city is online and monitored for better safety and security. It can pose potential privacy and security issues [12] [13].

Internet of things (IoT) is a crucial element for building a smart city. Petrolo et al. [14] discussed integrating cloud and the internet of things in a smart city. Privacy is a grave concern for people living in a smart city. Zoonen et al. [15] discussed various privacy issues, directions, and concerns about privacy issues. A framework has been proposed that measures these issues of smart city citizens. Arasteh et al. [16] represented the basic understanding of IoT technology and their potential usage in future smart cities. As IoT is one of the most efficient technologies for integrating various daily life services, it has its practical use in smart cities. Arroub et al. [17] included several smart city models and their applications. In this paper, the authors discussed the literature describing the trends and models of smart cities. Medina et al. [18] described IoT based smart city attributes and applications extensively. Trindade et al. [19] provided a detailed description of recent sustainability issues and their solutions. Similar to Medina et al. , Elrawy et al. [20] presented a brief overview of IoT based smart city security threats, protocols, intrusion issues, and countermeasures. A holistic survey of various attack detection systems has been put forth extensively by the author. There has been a significant amount of literature related to multiple aspects of the smart city. Ekai et al. [21] presented a brief description of the advantages, disadvantages, and smart city issues and their application.

Silva et al. [22] represented a brief overview of smart city architecture and a brief description of smart cities around the world. Al-Ani et al. [23] proposed a conceptual framework and applications of a smart city. Challenges and issues are also described in this paper briefly. Al-Smadi et al. [24] provided a comprehensive, precise overview of recent literature about smart city issues, including a comparative analysis among these. Curjon et al. [25] described the current situation of smart cities around the globe. The authors also tried to find out recent issues regarding smart city privacy and their state of the art solutions. Sánchez-Corcuera et al. [26] presented various dimensions of defining a smart city, including a brief description of smart city application sectors and research challenges. Chen et al. [27] provided a comprehensive survey of freshly growing research areas of smart city applications and artificial intelligence to provide potential readers with a solid understanding of the trends and open research issues. Lau et al. [28] and Ma et al. [29] described a similar artificial-intelligence-based approach for leveraging various issues of a smart city. Both of these literature focuses on the vast amount of data collected by multiple sensors and other IoT devices installed in numerous smart city infrastructures. Bhushan et al. provided a solid understanding of blockchain and smart city trends and their integration scenario. Blockchain is one of the recent technologies that can be used in many applications used in our daily life. Kirimtat et al. [30] discussed smart cities in one of the latest literature related to smart city surveys. In this survey, authors have extensively explored the literature related to smart city surveys and provided a useful overview of smart city applications, elements, and prospects.

Based on the literature discussed for this paper's feasibility and suitability, to the best of our findings and imagination, there is a need for a comprehensive review of the smart city architecture, application, requirements, security, privacy issues, and proposed solutions. This study will be a crucial source of information for potential scientists, researchers, and learners. A summary of this work can be described as follows –

- Benefits of a smart city for the future
- Smart city components and requirements in details
- Details and precise description of smart city requirements and architecture

- Brief description of various applications of a smart city, including their recent advancements.
- A detailed explanation of smart city security requirements and security issues
- Privacy issues of a smart city are discussed concisely along with state of the art issues and solutions.
- Future research directions that can lead to innovative solutions by potential scientists and enthusiasts

## 2. Smart City Background

Urbanization is becoming difficult due to the ever-increasing population. For this reason, education, transportation, energy, healthcare, etc. need proper management. Transparency, trustworthiness, optimization, and monitoring are required for better administrative decision making and implementation [31,32]. It is widely agreed that sustainability, decentralization, and information integration, and communication technology are crucial for smart cities [33]. The aims and objectives behind smart cities' inception are to improve citizens' socio-economic conditions, create a better work-life balance, produce and implement innovative sustainable solutions. The subsections below highlight the importance of a smart city, its components, pillars, requirements, and architecture.

### 2.1 Need for smart city

Smart City has the potential of being one of the next significant technological advancements. It incorporates technology in every aspect of life to improve QoL. Some notable contributions of smart cities are discussed below.

- Sustainability: Smart cities aim to build smart and sustainable healthcare, sustainable energy consumption framework [34], and inspire a greener lifestyle. Social issues, climate change, waste and pollution management, and using natural resources intelligently are crucial for building smart cities. Technological advancement often reduces and degrades the emotional and physical well-being of humans. Smart cities will put humans at the center and adjust technology to improve their QoL.
- Security: Smart city requires secure communication, monitoring, and response. Each element of a smart city is interconnected through various devices. These devices help connect citizens of the smart city also. Smart city infrastructure, state of the art technology improves physical architecture and cybersecurity situation. Blockchain and other technologies are used as a security provider for the smart city. It uses cryptocurrency and peers to peer communications. The pseudo-anonymity of blockchain technology secures the transmission of big data in the smart city. Blockchain creates an environment of trustless transactions that prevents fraud and eliminates a third party [35, 36].

- Connectivity: Smart city thrives on connectivity. A collective approach through an interconnected society for learning will increase creativity and ultimately contribute to making smart humans. Connectivity in a smart city is provided using both wired and wireless medium [37].

- Decentralization: One of the main goals of building smart cities is decentralizing governance, healthcare, education, finance, etc. So that people can receive services anytime, anywhere, without restrictions. If individual records (i.e., property ownership) are stored using blockchain technology, transparency and immutability shall prevent fraudulent activities. Digital identity is already being utilized in some counties. It can be used for authentication purposes at an individual and business level[38].

## 2.2 Smart city Components

ICT is very instrumental in smart cities because the information collected through sensors and other services-oriented applications is processed to create a better living and governing experience [39]. Wearable devices, smart surveillance, etc. collect data from the environment. All types of monitoring, decision, and operation are made based on this real-time data [40].

Communication technologies can be both wired and wireless. Therefore, it is necessary for any system to seamlessly switch between these networks and gather information from all of them [41]. The smart city's processing units process the collected data from the sensors and the collaborative heterogeneous systems. This information are synthesized and later used to predict the state of the city. The processing arena comprises cloud computing services, data centers, and other systems involved in the decision-making process [42]. After gathering and processing information from all aspects of city lives, the last step is to predict residents' needs and wants. This helps the government and other organizing bodies to improve the quality of lives and services better in the city [43].

## 2.3 Smart city Pillars

It is proposed that the four pillars of a smart city consist of the physical structure of the city, the institutions, the society, and the economy [44]. These pillars of smart cities are briefly discussed below.

### 2.3.1 Physical Infrastructure

It is one of the most important of all the infrastructures. The quality of life depends on it. The architecture and infrastructure include smart and green buildings, smart vehicles, IoT integration, smart grid, etc. Urban areas should be planned and implemented to inspire maximum resource utilization [45].

**2.3.2 Institutional Infrastructure:** The government of a smart city should be all-inclusive and transparent. There should be harmony between the citizens and the administrative body of the city. All types of public, private, and hybrid institutions are included here. They should run the city with a proper governing body. The organizations will have to maximize the use of human resources that will create a sustainable society. Solutions will come out effectively from the participants of the smart city. As urbanization is heavily dependent on technology, modern technology can be another impactful addition to infrastructure enhancement [46].

**2.3.3 Social Infrastructure:** Smart society can only be built if the people of the cities are smart. The collective intellect of citizens should be encouraged to develop in such a way so that human capital can be maximized. Sustainable society and quality of life can be improved through maximizing social awareness, responsible citizen building, and maximizing human resources. Connection and correlation among other administration institutions can go a long way for building social infrastructure also. Each person must have access to information and resources to manage their lives effectively. As social infrastructure is related to the betterment of social and industrial behavior and the people's relationship, this is one of the essential infrastructures of smart city building blocks. So smart, educated, and knowledgeable citizens of the smart city are very crucial for enhancing the social infrastructure [47].

**2.3.4 Economic infrastructure:** The economy must grow steadily to accommodate the increase of people and activities. Estonia has used blockchain technology to allow E-residency to people all over the world. It does not permit the E-residents to physically live in the country but will enable them to conduct business and own properties. It can also inspire building smart cities as to how the city can keep the economy flourishing without exhausting its physical space and natural resources. Ecommerce is a very crucial sector for the economic development of smart cities. Good e-commerce can range beyond the city and potentially expand the business. It will enhance the economic and financial growth of the society. Modern technology can be integrated with industry for efficient and effective automation. Thus, integration with other smart cities' infrastructure and utilizing natural resources using modern technology is crucial for economic infrastructure development [48].

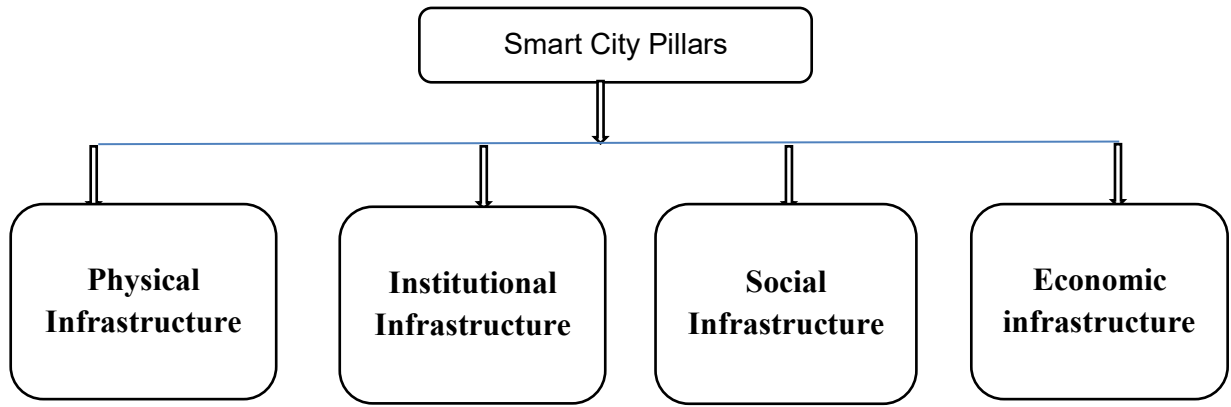

Figure 1. Smart City Pillars

## 2.4 Requirements of Smart City

Smart cities are the newest concept of urbanization. A smart city is interconnected, intelligent, and sustainable. Many sectors need to be taken into account to create a safe and secure smart city. These are as follows-

### 2.4.1 Smart Human Capital

Smart human capital means effective, efficient, and knowledgeable human resources. Human resources are one of the most critical factors for any kind of society. Citizens use their education, knowledge, creativity, and intelligence to develop themselves. Social learning is another factor that is immensely impacted by sustainable social infrastructures. Critical thinking is a crucial element for the personal and social improvement of a citizen. Proper monitoring and authoritative control over citizen's data are needed. As people inside a smart city are connected dynamically, a digital identity is required. Digital identity shall help the citizens to use all types of services seamlessly. Protecting emotional and physical stability is very important for the integrity of a smart city. Above all, ensuring the utilization of resources smart human capital is crucial [49].

### 2.4.2 Smart Infrastructure

Developing the infrastructure of a city is vital for its development. Smart buildings are infrastructural improvements in smart cities[50]. The smart building has efficient energy consumption and environmental protection. Buildings can use solar energy. Various sensors and IoT technologies are used in such an infrastructure for better performance [51][52][53]. The infrastructure development of smart cities also includes smart transportation systems. Driverless and hassle-free transportation is envisioned for smart city infrastructure. Various types of sensors are used in roads and vehicles to collect data. This data is collected and analyzed for proper decision making. Global positioning systems and different vehicular ad-hoc network infrastructures are also immensely crucial for smart infrastructure. Besides, smart grid technology [54] is implemented for efficient energy consumption. This shall provide citizens a better quality of life.

Li [55] proposed a methodology of implementing blockchain in smart city infrastructure. The author analyzed big data-enabled infrastructures and issues. IoT, a crucial part of the infrastructure, is related to every aspect of a smart city. Tanwar et al. [56] provided a detailed explanation of IoT based smart city, smart grid, and their importance in smart city infrastructure. ICT is a driving factor for smart city urbanization. The transportation, energy sector are also crucial. Mwaniki [57] discussed elaborately the recent development of Nairobi as a potential smart city. Nairobi can be an example of an ICT based smart city framework. Kumar et al. [58] discussed a similar approach. The authors discussed physical and ICT infrastructure and the other two potential vital aspects that can significantly impact smart infrastructure transformation development. Caird et al. [59] analyzed design aspects of various smart cities of the UK. This analysis provides an essential insight into both infrastructure, communication, and other elements.

### 2.4.3 Smart Services

Various social, financial, and municipal services are crucial requirements in smart city services. Health care service [60] is one of the vital benefits of a smart city. This service includes a patient record management system, intelligent appointment system, data storage, and sharing among the related organizations. IoT and sensor devices can be used for patient data collection. Remote healthcare services can be facilitated for the persons who need attention. Moreover, healthcare data can be analyzed for disease detection and medication. Financial services like banking, insurance are also very crucial for citizens [61]. City administrators should be able to dispatch essential services whenever anyone needs it. Online banking is already a growing sector. Smart insurance services can also be implemented for facilitating a state of the art services to the citizens [62]. Besides, education is one of the crucial demands that need to be fulfilled by the administrator. For this reason, efficient and smart education is also a vital requirement for smart cities. The learners need to be able to learn and use their knowledge in real-world scenarios for sustainability.

### 2.4.4 Security and Privacy

Modern urbanization is widely dependent on sensors and IoT devices. These devices are used in various infrastructures. Data collected from different devices is analyzed for better strategic management. Administrative decisions and procedures can be improved significantly using data analysis. A security breach in any of the infrastructure can potentially jeopardize the whole system. People of smart cities are always monitored for social security purposes. As smart cities thrive on collecting and processing citizens' personal, locational, and sensitive information, any breach of privacy is threatening individuals and society.

Wearable devices and health monitoring equipment collect physical information. Wearable devices collect a person's location, heartbeat, etc. healthcare sensors and devices collect patients' health-related data. For both cases, the data is personal and should not be an open or public view. Breach of privacy can leave people vulnerable to severe attacks [63]. Smart cities are heavily based on locations. Intelligent traffic, intelligent and autonomous vehicles continuously transmit locations and vehicle information. This helps with emergencies, maintenance, and insurance. If this data is not protected, criminals and stalkers might commit irrevocable damage. Inherently connected social life improves the psychological state of citizens. This helps with smart human capital. Moreover, constant monitoring of citizens for security purposes is also needed. So, proper privacy and security measures for a person's digital identity and footprint are a much-needed requirement for a smart city [64].

Khatoun et al. described a holistic overview of smart city design specifications, cybersecurity issues, and recommended solutions elaborately. The smart city being the apex of digital cities, the interconnected nature creates opportunities for various digital crimes. Besides, privacy and security issues are already there. Losavio et al. [65] described these issues extensively. E-governance is mostly dependent on ICT. The decision making, process management, control, etc. creates several privacy issues. Yang et al. [66] addressed those issues and proposed a framework focusing on the essential privacy and security requirement for a smart city. IoT security is crucial for smart city infrastructure security [67]. Butt et al. [68] discussed IoT based smart city infrastructure, data acquisition, and processing from those infrastructure and security issues.

Khatoon et al. [69] described healthcare-related security issues, i.e., patient data security, connectivity security, and privacy, along with their solutions extensively. Badii et al. [70] conducted a detailed study describing GDPR oriented IoT infrastructure for a smart city. This study discusses devices and application-related data collection, processing, and management. Tubaishat et al. [71] followed a case study-oriented approach. In this study, the authors have briefly described the big data perspective. A holistic framework of UAE smart city is analyzed and studied for understanding smart city security and privacy perspective.

### 2.4.5 Quality of life and Sustainable Environment

All technology provides a quality environment for living a healthy life. The government is service-based. Smart governance can be beneficial for increasing the quality of life. Administrative procedures and decisions should be automated and transparent to the people. Coherent correlation among other smart city infrastructures components shall create harmony, and citizens will enjoy standard benefits. Smart cities must be sustainable and environmentally friendly. Rapid urbanization, the rising number of industries, and diminishing greenery quicken climate change. Therefore, creating the scope of living a greener lifestyle is mandatory. These ideas will be taken into account while planning smart cities. Smart Grid is an excellent solution for irresponsible energy consumption. The usage of renewable, shareable energy can help smart cities to reduce carbon footprints. Proper waste management to reduce pollution and biodegradable materials is crucial for a cleaner and greener environment [72].

### 2.5 Smart City Architecture

Smart cities are a relatively new idea of urbanization. Though several aspects and features of the smart city are implemented, a holistic implementation of a smart city is yet to be implemented. Researchers and scientists are trying to bring forth a real-life architectural pattern of a smart city. Defining a definitive architectural model is essential for any kind of technology. Without a proper design, the technology will not have any standard to follow, and there is a significantly less chance to evolve. A smart city is highly dependent on data since all decision-making is done after a proper analysis of the collected data from various infrastructure. Considering the proposed works and approaches, four layers have been identified. A brief description of the smart city architectural layers are described as follows.

### 2.5.1 Sensing layer

Most of the decision-making process is dependent on efficient data analysis. For this reason, proper data collection is a very crucial task. On the other hand, data collection from various infrastructures is a very daunting task [73]. Smart city infrastructure comprises various elements such as sensor networks, IoT devices, smart grids, etc. Sensors collect data from different types of installments. Temperature, humidity, air pressure, etc. are prevalent for weather forecasting detection. These are used inside smart buildings and small weather stations also. The sensor network is also used in parking. IoT enabled devices and applications are very crucial for healthcare, medical, intelligent transportation, etc. So, sensors, including IoT devices used in various appliances, are significant data collection sources. Some of the sensors that are used to collect data are RFID, Infrared Sensor, Temperature Sensor, Bluetooth, humidity sensors, cameras, etc. Sensors collect data with different matrices [74][75]. Since data collection helps with decision making and automation, increased data collection vectors shall enhance a genuinely smart city[76].

### 2.5.2 Transmission Layer

In the previous layer, a network of devices connected collects data. After the data collection, it needs to be sent to the destination. Besides, the connected devices also need to communicate with others. Communication among the devices helps make decision making, fault detection, remote maintenance, etc. too. For example, a wireless sensor network can facilitate remote maintenance in a few instances. The communication technology used for data processing (send/receive) belongs to the transmission layer. This is compared to the backbone of the total city architecture. Therefore it is crucial to use efficient data transmission technology considering the usage vectors [77, 78].

Communication technology can be divided into two primary modes. One is short-range communication, suitable for various low powered sensor networks [79] [80]. Long-range communication happens among typical smart devices such as mobile devices and different wired and smart wireless devices. 4G/LTE is the recent addition of cellular networks with wider bandwidth and faster transmission rate. WiFi is a prevalent and widely used wireless local area network. Its range is lower than cellular networks and generally can work on mobile infrastructure. Throughout time, both shorter and broader range of data communication technology has been developed and augmented. As the need for communication has increased, the sensing layer's device specification and types have also increased [81] [82].

### 2.5.3 Data Management Layer

This layer's functions include data processing and management. Data analysis is one of the most important attributes as well as functionalities of sustainable smart cities. Various infrastructure produces data. Citizens also generate large amounts of data every day [83]. Confidentiality, integrity, availability, and authenticity are the basic properties of data that need to be ensured. Moreover, data from various sources need to be processed, such as cleansed, edited, and merged for future usage. As the amount of data is vast, data storage and processing raise a concern. Moreover, as time passes, infrastructure and services will increase. So, the data storage should be scalable also. In this case, data centers need to be built for data storage and backup [84]. A variety of data sources results in data heterogeneity. It will improve data quality to a great extent. A smart city requires real-time data analysis in case of traffic management, disaster management, etc. To facilitate this purpose, advanced data analysis algorithms need to be also implemented. Effective and enhanced data storage and analysis can provide smart and real-time decision making for smart cities.

### 2.5.4 Application Layer

The residents of smart cities enjoy various features of smart cities through this layer. Hence, this layer has immense importance as a service provider to the user. Data analyzed from the previous layer is represented here as a decision. The people are not aware of the underlying working principle or algorithms used as the brain of a smart city; rather, they want the end result. This result comes as smart healthcare and medical technology, smart energy, smart waste management, smart agriculture, smart education, etc. Smart healthcare provides state of the art medical facilities, including healthcare data management, disease detection, 24/7 medical service, etc. These facilities shall impact a person's personal and professional life.

Smart grid shall ensure load balancing, smart and uninterrupted electricity availability, efficient distribution technique around the city, automated electricity consumption statistics, bill generation, bill pay, etc. Several other application services will discuss details in the next sections. As this layer directly affects the smart city residents, this is also prone to various security threats. Any attack in this layer shall hamper the infrastructure and the people. Moreover, this layer should also deploy scalable applications as the users will increase with time [85].

Table 1. Smart City Layers and their Features

| Attributes | Sensing layer | Transmission Layer | Data Management Layer | Application Layer |
|---|---|---|---|---|
| Functions | Collecting data; Interaction with environment | Communication among connected devices; Data Transmission; Fault detection; Remote maintenance; | Data processing; Data Analysis & Management; Protecting Confidentiality, Availability, | Providing uninterrupted services to the user; Interacting with the application user; Data Visualization |

| | | | Authenticity, Integrity | |
|---|---|---|---|---|
| Devices | Various sensors, Other IoT Devices, | Transmission Devices | Storage Devices; Data processing equipments | Smart devices, Smart distribution, monitoring, tracking devices. |
| Connected layers | Physical world; Transmission Layer. | Sensing layer; Data Management Layer | Transmission layer; Application Layer | Data Management Layer |
| Technology | Temperature Sensors, humidity sensors, RFID, Bluetooth etc. | 4G/LTE, WiFi, WiMaX, other short and Wide Range Communication | Data Processing, Analysis technologies, Decision-making algorithms | Application Development, Smart Grid, Healthcare Monitoring, Home automation etc. |

## 3. Applications of Smart City

A smart city comprises various applications. All the applications [86] of smart cities are for the betterment of the quality of life. The residents shall be able to enjoy the benefit of modern technology. Different applications of smart city are described below:

### 3.1 Smart Transportation System

There are various modes of transportation, such as Air transportation, Water transportation, and Road transportation. Transportation is the way people communicate, and businesses supply logistics. Vehicular ad hoc networks are the newest addition to intelligent and autonomous transportation. This facilitates inter-vehicular communication as well as vehicular internet connectivity [87]. Vehicles are connected to each other and with the network infrastructure alongside the road. Rather than the conventional transportation system, vehicular connectivity facilitates real-time traffic analysis. This intelligent traffic system provides seamless connectivity. Traffic analysis helps congestion control, handling emergencies, detecting illegal activities on the road, and pinpoint the object of interest. Potential traffic analysis for airline industries that facilitates smooth, safe, and flexible air travel. Various sensor networks collect weather data, analyze it, and send the air traffic controller's decision. The same data is also used on onboard computers for intelligent decision making [88]. Real-Time radar tracking is also possible for each aircraft in the sky. The same goes for traveling using waterways. Various commercial and personal vessels can be tracked in real-time anywhere around the ocean. In this case, GPS technology is used. The data is sent using the API in real-time. The same can be applied in smart city water travel using rivers and canals.

Transportation is a fundamental need for smart city citizens. Transportation schemes of different mid-sized cities of the USA were compared according to the indicators mentioned above to find their difference. Giang et al. [89] analyzed a fog computing-based transportation technique. Smart transportation techniques include intelligent vehicular networks, smart traffic infrastructure, etc. These methods pose a challenging scenario if applied to a fog computing environment. Giang et al. briefly describe this. Big data analytics can help leverage smart transportation problems. Data collected from transportation infrastructure can improve the shortcomings. Khazaei et al. [90] described such an approach in Toronto, Canada. Shukla et al. [91] proposed real-time analysis can leverage congestion control for smart transportation.

Similarly, another big data-oriented approach to smart transportation. The authors of the paper aimed for better service, a greener environment, and sustainability. Kelly et al. [92] reviewed and analyzed smart transportation schemes of various cities reviewed and analyzed by Vehicular technology advancements, traffic systems, scope, and geographical scope. A smart city comprises IoT devices. These devices are connected via ICT and developed using various techniques. Yan et al. [93] proposed a framework combining these techniques. This study also uses Chinese trans+portation for framework evaluation. Boukerche et al. [94] addressed traffic control and crowd management solutions unexplored in smart transportation infrastructure. The authors also mention various challenges, issues, and possible solutions to congestion control. Lin et al. [95] designed a securing transportation infrastructure using drivers' location data securing using a homomorphic approach. In traditional scheme drivers, the authors proposed that the location collection technique reveals privacy issues; hence, the location is encrypted and stored in cloud services. The transportation system's performance can be intensified using the data collected by various devices used in the system. A machine learning-based approach towards smarter and efficient transportation can be a crucial addition to smart city infrastructures. Zntalis et al. [96] combined all the studies related to this technique and presented a concise and solution-oriented analysis.

## 3.2 Smart Healthcare System

Healthcare is a basic human need. The number of patients is increasing with diverse diseases. The traditional healthcare system needs to be improved with modern technologies. Expert doctors are not rising concerning the number of patients. Hospitals also have to accommodate this situation. This delicate situation is addressed with various sensors, IoT devices, and intelligent disease detection techniques. Wrong medicine prescription and inappropriate disease detection can cause severe health damage to the patient. THE recent COVID-19 epidemic has indicated that our healthcare system needs to be significantly improved to accommodate epidemic cases. An automated and intelligent healthcare system can be a suitable solution for this scenario. Sensor enabled IoT devices are useful to measure a patient's condition. Sensors can measure patients' heart rate, blood pressure, oxygen saturation level, etc. These attributes are very crucial for pre and post-diagnosis. Pacemaker devices are installed for constant monitoring of heart patients. Patients' historical data can help an efficient diagnosis. All of these technologies are integrated with smart and secure hospitals. Sensors and paging devices can be used to detect doctors' and nurses' location. It will help with emergency medical team dispatch [97, 98].

Smart medical and healthcare technologies will produce a significant amount of data. The patient's data taken while admitting or prescribing, sensor-based data, health records, and prescriptions are also stored. These are very sensitive personal data for patients and need an enhanced data security and privacy. For this reason, organization data privacy, protection, and security policy can be implemented [99, 100]. To leverage issues recently, several blockchain-based healthcare data management systems are proposed. Since blockchain technology is a growing idea for data security and privacy, it can be a suitable solution for this purpose [101, 102]. To protect the human capital of a smart city, healthcare systems need to be compatible with demand and supply.

Healthcare technology has advanced recently within a few years. Adult healthcare is important, especially on a community basis. Tomita et al. [103] proposed an integral approach of smart home and healthcare for elderly citizens of the community. This approach addresses the overall healthcare facilities for elderly people staying at home. Aziz et al. [104] proposed a sensor-enabled approach for monitoring health conditions remotely. Wireless sensors will measure blood pressure, temperature, and send the data remotely. The concerned doctors and medical teams can then analyze the data for possible disease prevention. If the collected data passes the alarming level, the doctor will be notified instantly. Chui et al. [105] analyzed the healthcare researches in disease detection in their paper. The authors analyzed the artificial intelligence-based approach for cardio and other disease detection that provides an essential insight into future research direction. Mahmoud et al. [106] provided cloud and the IoT related approach for healthcare systems. The authors analyzed the possible issues of sensitive health care devices and their standards. They have successfully identified potential research opportunities for energy-efficient healthcare infrastructure.

Similarly, Chen et al. [107] proposed an edge computing-based approach. The authors have found out that, edge computing approach for the healthcare system facilitates flexibility and optimization of the user experience and resource utilization. The electronic health record management system is the newest addition to the modern healthcare system. As this record contains patients' data and sensitive health information, better security is a crucial need. Zhang et al. [108] and Tripathi et al. [109] provide a very efficient approach for securing healthcare data using blockchain technology. The authors of both papers emphasized addressing the current challenges for securing health data. These two pieces of the literature indicated some potential research direction for the healthcare system development Tanwar et al. [110] proposed a similar approach. The authors propose a blockchain-based approach to healthcare system optimization. Performance analysis of healthcare framework integration in various blockchain platforms, process management, healthcare record management system, and improving the overall latency is discussed and analyzed by the authors. Ahad et al. [111] examined the recent advancement of healthcare data communication. The authors emphasized the 5G approach for reliable, integral, dedicated communication techniques in healthcare data exchange among various applications, users, and medical personnel.

Ismail et al. [112] proposed a reliable, secure, privacy-preserving, minimal network traffic enabled healthcare framework based on blockchain technology. The authors also analyzed robust transactions in communication over the network, which increases the throughput. The framework is a very lightweight framework that reduces overhead resource consumption and addresses blockchain-based healthcare technology's current issues. Similarly, Abou-Nassar et al. [113] proposed another blockchain-based model for reliable and sustainable healthcare infrastructure. This model uses a private blockchain for application purposes. Access to healthcare facilities needs a proper distribution of resources throughout the region. A smart city demands an efficient supply of healthcare facilities and the necessary amenities for the citizens. To make the total system efficient, Oueida et al. [114] proposed a model that uses stakeholder satisfaction of healthcare facilities. Artificial intelligence is used for disease detection. Ali et al. [115] provided an in-depth learning approach for accurate disease detection and monitoring systems. The proposed model is efficient in feature extraction from the healthcare-related sensor and medical data and analyzing those for better prediction.

### 3.3 Smart Power and Energy Management System

Energy is the driving factor of any infrastructure. It powers up the buildings, home, industry, hospitals, schools, colleges, and other institutions. Hence, smart and efficient power management systems are essential for a sustainable smart city. Renewable and nonrenewable energy sources are the two types of energy source types. Nonrenewable energy sources need to be used efficiently at present so that it can be preserved for the future. This will make the nonrenewable energy source sustainable. Coal, gas, fossil-fuel, etc. are some of these types of energy sources. Traditionally these are the types used as power sources. As the amount is limited, for this reason, there is a need for smart and sustainable solutions [116] [117]. Smart energy management refers to a balance among demand, production, and supply of power. Besides, it incorporates other energy sources, e.g., renewable energy.

Renewable energy is a relatively new addition as an energy source. The use is increasing day by day rapidly. Nowadays, solar energy is used at home, also [118]. Some of the smart buildings use renewable energy as an energy source. This reduces environmental pollution by reducing carbon and other harmful materials being released in the air. This increasing demand for renewable energy is a source of micro-grid [119]. Hybrid power generation can be another source of energy that harmonizes two or more power sources. This emphasizes a sustainable environment in the energy sector. A smart energy system shall comprise easy and flexible accessibility of power for all citizens, optimized supply of electricity, use of renewable energy systems, and a combination of all types of resources, including the source and distribution. If all the citizens can have power at their home, the usage statistics can be used for optimized power distribution in various cities. This technique shall ultimately facilitate an economically feasible solution.

A smart city promotes an echo-friendly, greener, and sustainable environment. The electric vehicle can be integrated into the smart grid so that. Tan et al. [120] described the framework for vehicle integration in the smart grid, possible issues, and future research directions. Big data can provide useful insights as it contains data from various sources and contains versatile features. Smart grids also regularly produce vast amounts of data collected from sensors, IoT devices, communication channels, network traffic, user feedback, usage statistics, etc. Zhou et al. [121] proposed the usage of big data collected from smart gids for better energy management, distribution, governance, and production. The smart grid can be sustainable and secure if the analysis is done constructively. Rivera et al. [122] provided a cloud-based approach to smart grid analytics and grid management. The authors also proposed service level agreements for this purpose. Olabi [123] described various energy storage systems. Storage technique classification and storage selection criteria are described briefly and concisely. Mengelkamp et al. [124] adopted a decentralized approach to address smart energy. The authors focused on local energy markets. This approach uses blockchain technology and a decentralized market concept. The authors analyzed the economic feasibility of the concept for implementation purposes. Energy management requires proper decision making and governance. The energy sector produces a huge amount of data. This data can be analyzed for proper decision-making purposes. Marinakis et al. [125] proposed a framework that is based on big data for energy sector management. This big data-oriented intelligent grid management approach facilitates useful visualization for various stakeholders of a smart city. A smart energy management approach can be intensified using reinforcement learning algorithms. Kim et al. [126] proposed an intelligent energy management framework for smart buildings. This approach can provide an optimized approach in case of energy use and cost-efficiency based on learning algorithms. Abate et al. [127] provided an IoT enabled approach for smart grid technology through smart metering. This IoT enabled smart meter uses an algorithmic approach for electricity consumption monitoring, reliable and accurate communication. Security and privacy of the smart grid are very crucial. Gai et al. [128] proposed an approach for securing the smart grid. The authors proposed approach aims at stakeholder's privacy and smart grid security.

## 3.4 Smart Network Connectivity infrastructure

Connectivity is the backbone of any infrastructure today. The same goes for smart cities. The elements of a smart city need to be connected. A reliable, trustworthy, and the dedicated connection is required for data transmission among the elements. It helps make a system context-aware, robust, and efficient. Here, connectivity can be both wired and wireless. Almost every resident in today's urbanized work uses a smartphone. The cellular network facilitates 3G, 4G, and LTE connectivity. It enables seamless voice and video communication over smartphone networks. The most widely used wireless LAN network includes WiFi, WiMAX, etc. This connectivity infrastructure is included in almost every corporate, home, industry, educational institution, and other networks. The connectivity helps people to access services and data from anywhere in the world. WiFi is also one of the most affordable wireless network installations. Wired connectivity using broadband and optical fiber connection exists for more reliable and dedicated connections [129].

Wireless sensor networks (WSNs) are one of the most widely used connectivity for IoT based networks [130]. Sensors devices are used in various infrastructures of smart cities. WSN is used in weather stations, waste management, industry, home automation, etc. Another important network is the vehicular ad-hoc network. This infrastructure is used in intelligent traffic systems. Satellite communication is used in smart cities for reliable communication for various satellite TV channels, Military communication, Weather status update, and emergency communication [131]. All of this infrastructure is connected to build interconnects of network devices. All the devices help the residents of the city to enjoy different types of application services.

The connected infrastructure has storage architecture in the cloud as well as local storage devices. Cloud services have become very popular and useful nowadays. The services can be provisioned within the shortest possible time and easy to use. There is no need for extra infrastructure to be installed for this purpose [132] [133]. Network infrastructure has to act smartly by load balancing, automatic network metering, malicious traffic detection, threat analysis, network backup provisioning, remote maintenance and emergency decision making, etc. An intelligent intrusion detection system can be implemented for any network threat detection [134,135].

## 3.5 Smart Home

A smart home [136] is modern infrastructure in the modern world. The smart home concept is used even before the idea of a smart city. Scientists and researchers are always trying to make domestic life comfortable. A smart home is based on sensors, IoT devices, GPS, alarm systems, dedicated network connections, etc. Solar energy is integrated into smart homes now.

For this reason, energy management systems are also included. Energy efficiency is ensured in smart homes by using smart lights, ambient sensors, and weather monitoring. Electricity consumption data is always collected from various devices. This data can be analyzed, which can result in proper electricity distribution in multiple devices. The same can be done using a prefixed budget. Unused devices can be turned off if not used frequently.

A smart city is incomplete without a smart home. The citizens need to feel safe while staying at their home also. For this reason, a constant monitoring and emergency system needs to be installed. Emergency systems shall include quick, reliable, 24/7 communication with hospitals and police stations. So that emergency healthcare and emergency services are ensured [137] [138]. As a smart home is connected to the internet like any other infrastructure, the data produced by this infrastructure needs to be stored and processed in a secure and trustworthy environment. A smart home is a crucial part of ensuring the quality of life for the residents. For this reason, this infrastructure needs to be improved with time according to user demand.

Sourantha et al. [139] proposed improving the smart system's security system and automatic system. The proposed method approach uses movement detection using PIR sensors and later uses raspberry pi for capturing the image to detect the object. Liu et al. [140] proposed a security system related to cybersecurity. This approach detects cyber-attacks associated with the increment of pricing of the smart homes. The attacker illicitly increases the smart home bill of other persons whereas significantly decreases his own bill. Khan et al. [141] described an energy utilization approach for smart homes. This IoT based approach uses Zigbee networking for communicating among the devices and controls the energy usage inside the smart home. The authors proposed an intelligent system for controlling energy consumption, which is not affected by any external entity. Kang et al. [142] proposed smart home security related advanced framework. This security framework enhances security, integrity, and access control of smart homes' smart devices. Malche et al. [143] proposed an IoT enabled smart home system that emphasizes control and automation. Naik et al. [144] provided a similar IoT based approach but an open-source framework. This open-source increases the communication reliability and security features of smart homes. The quality of living in a smart home is a concern for the researchers. Feng et al. [145] proposed an interactive, intelligent, reliable, and user-friendly smart home architecture that specifically aims to develop the living environment. Al-Kuwari et al. [146] and Singh et al. [147] proposed IoT based smart home controlling, monitoring, and automation approaches using sensors.

Securing a smart home is crucial for the smart city and providing user data privacy. Zhang et al. [148] provided a privacy-oriented secure framework for smart homes. This proposed approach uses a fog computing-based approach smart home model. Gajewski et al. [149] provided an architecture for smart home cyber-attack detection and possible countermeasure. Similarly, Ferraris et al. [150] described a privacy-preserving approach to build trust models. This model addresses security threats and privacy issues. The authors briefly described the threat models for smart home, potential unauthorized access control management issues, and feasible solutions. Guhr et al. [151] described smart home user data security, privacy issues related to healthcare. The authors discuss the possible behavioral, psychological concerns related to smart home data security, device security and addresses the problems with proper recommendations.

## 3.6 Smart Office

Offices need to be made sustainable and smart for better productivity. Employees need to be provided with all types of facilities for their efficiency [152]. The buildings or infrastructure of an office is equipped with modern amenities, including all sorts of technology. Proper controlling and automation are also installed for temperature controlling, air conditioning, smart light, intelligent security system, state of the art connectivity, etc. Sensors are used to measure the temperature and humidity so that the airflow and humidity can be controlled. These devices are also used for detecting any human presence in an office room. This will help regulate electricity in proper places, rather than wasting it. The same goes for air conditioning too. Automation happens through real-time data analysis. The data is collected from various sensors and IoT devices. The data is then analyzed for decision making [153].

Proper monitoring is also in place. All the employees and staff are always monitored as long as they are in office. RFID based identity cards are issued for entry and exit purposes [154]. This ID card is also used for authorization purposes. The access right is provided against their identity card. This technique will prevent unauthorized access and entry. Moreover, proper tracking due to security purposes can be implemented using this card. Apart from this technique, a more efficient fingerprint or iris scanner can also be used for enhanced security. Office buildings can use renewable energy, which is environment friendly as well as efficient [155]. Similar to other infrastructures, proper data protection and privacy mechanisms need to be implemented in smart offices.

### 3.7 Smart Identity Management

Proper identity management is essential for efficient service providing to every citizen. Smart citizens need a smart identity. Smart city implements an identity management system that will act as a unique identifier for all types of tasks such as opening a bank account, paying taxes, admission to educational institutions, hospitals, applying for jobs, etc. This shall facilitate the citizens with a one-stop service for availing the necessary services against the identification number. Essential personal information can be stored against this identity number as a starter. Later, as the work vector increases, the data will also be increased. Private insurance, health insurance, everything can be issued against this one number. This shall facilitate the administrator with enhanced information management. Identity theft can also be reduced in this case, as one has to have all the information of a person's lifetime to conduct identity theft [156].

As it will facilitate the city administrator and users with a lot of advantages, the risk shall also increase. This is potentially vulnerable to universal points of vulnerability. Any compromise with this identity can harm a person to a far greater extent. For this reason, proper data encryption, along with another state of the art data security and privacy measures, needs to be implemented. Data can be stored in separate databases. Blockchain-based identity management systems can be used for identity management. It will ensure data immutability, authenticity, and integrity. Besides, it can be accessed from anywhere in the world securely. Integration with other services is also possible if a universally blockchain-based system is used [157, 158].

### 3.8 Smart Administration

Citizen centric and service-oriented smart decision making is an impactful factor for a sustainable smart city. Modern technological tools and interfaces can be used for efficient data analysis. Behavioral data and usage statistics can be used for service modification and improvement. Citizens' opinions and feedback can be taken into account for a better quality of living. Smart city administration needs to provide the citizens with proper opportunities and a platform for expressing their ideas and views. They should be encouraged to make creative, smart, and sustainable decisions for the betterment of society. Participation in various operating bodies' vital selection and election procedure can inspire the residents and make them realize as an integral part of the process [159] [160].

Various administrative decision-making processes need to be quick and timely. Paperless managerial procedures can be introduced to make the whole process faster. Digital record-keeping of every task will be more comfortable. The electronic voting system is a very effective way of maintaining transparency in the election process. A blockchain-based voting system is also another secure and trustworthy voting mechanism that can be used. Another important task of the administration is to pay attention to what the citizens want. They will be able to send feedback or requests. The entire process can be displayed to the user's end through a tracking system. This system shall contain comments on the suggestions and feedback. A suitable administration can protect citizens' rights and keep the balance of expectation and reality in a smart city [161].

### 3.9 Quality Education System

Protecting sustainability, reshaping the smart city, and its improvement needs quality human resources. Education is the best way to produce smart and creative human resources. Modern ICT, integrated with the education system, has created some smart and useful ventures like e-learning through online courses and training. An intelligent design can be applied to the education system for its improvements. Students' psychological conditions and behaviors can be used for detecting potential gaps in the existing education system. Course materials and lectures are stored online for better learning and teaching. The online examination also takes place in case of emergencies and for overseas courses. The student identity management system facilitates access management, attendance, result query, etc. [162].

An online degree verification system is already in place. This is used by recruiters and admission authorities of various educational institutions. Students' records are stored digitally. This record is accessible from anywhere in the world online. Degree verification systems can be integrated into the human resource section of recruitment organizations for efficient recruitment purposes. Besides, there are other document verification systems. The integration of educational institutions with other infrastructure can create a suitable environment for producing smart human capital [163] [164]. Applications of a smart city can be summarized below, including their characteristics.

Table 2: Summary of Various Smart City Applications (Smart Transportation)

| References | Year | Contribution |
|---|---|---|
| Giang et al. [89] | 2016 | Addressing limitations of cloud computing-based vehicular network infrastructure; Vehicular network deployment using Fog computing; Analyzing the challenges involved in Fog computing-based architecture. |
| Khazaei et al. [90] | 2016 | A reliable, scalable, adaptive platform for traffic pattern recognition in the city area.; Congestion control, cruise control; |
| Shukla et al. [91] | 2016 | Data analysis for transportation optimization; Traffic pattern, congestion control, prediction, etc. |
| Kelly et al. [92] | 2018 | Vehicular technology adaptation in accordance with the population; Smart city vehicular technology topology adoption in various cities. |

| Zntalis et. al. [96] | 2019 | Using machine learning for transportation efficiency in smart cities. |
|---|---|---|
| Boukerche et. al. [94] | 2019 | Traffic control; Crowd management; Addressing challenges and issues for both the tasks; |
| Lin et.a l. [95] | 2019 | Secure smart transportation; Drivers location tracking; Privacy-preserving system for secure transportation. |
| Yan et al. [93] | 2020 | Combination of ICT and IoT for smart transportation framework design. |

Table 3: Summary of Various Smart City Applications (Smart Healthcare)

| References | Year | Contribution |
|---|---|---|
| Aziz et al.[104] | 2016 | Smart healthcare with monitoring, tracking using GPS/GSM; Patient-centric sensor-based healthcare architecture. |
| Chui et al. [105] | 2017 | Artificial intelligence-based approach for disease detection |
| Mahmoud et. al. [106] | 2018 | IoT based approach; Energy-efficient approach; Health-related device standardization and possible issues; |
| Zhang et. al [108] | 2018 | Blockchain oriented approach; Healthcare requirement analysis; Architecture design based on HL7; |
| Oneida et. al. [109] | 2018 | Artificial intelligence-based disease detection. |
| Ahad et. al. [110] | 2019 | Cellular technology-based approach for healthcare communication; Reliable and dedicated healthcare network for healthcare data integrity. |
| Ismail et. al. [111] | 2019 | Privacy-preserving approach; Proposed a lightweight framework for low traffic generated healthcare communication network. |

| Tanwar et. al. [112] | 2020 | Addressing current healthcare limitations; Proposed algorithm for data management; Efficiency analysis and smart healthcare tools and frameworks; |
|---|---|---|
| Abou-Nassar et. al. [113] | 2020 | Blockchain-based healthcare resource distribution across smart cities. |
| Tripathi et. al. [114] | 2020 | Smart healthcare using blockchain; Security and integrity ensuring healthcare framework; Potential issues and research challenges; |
| Ali et. al. [115] | 2020 | Disease detection based on deep learning; Feature extraction from health sensor and medical data. |

Table 4: Summary of Various Smart City Applications (Smart Energy Management)

| References | Year | Contribution |
|---|---|---|
| Tan et al. [120] | 2016 | Vehicle integration in smart grid and issues |
| Zhou et al [121] | 2016 | Big Data Analytics for energy management |
| Rivera et al. [122] | 2016 | Cloud service integration for smart grid management |
| Olabi [123] | 2107 | Energy Storage management and selection criteria |
| Mengelkamp et al. [124] | 2018 | Decentralized local energy market using blockchain |
| Marinakis et al. [125] | 2018 | Using big data for intelligent grid management |
| Kim et al. [126] | 2018 | Efficient and intelligent energy management for smart buildings |
| Abate et al. [127] | 2019 | IoT enabled algorithmic approach for energy consumption and management |
| Gai et al. [128] | 2019 | Smart grid security and privacy for stakeholder |

Table 5: Summary of Various Smart City Applications (Smart Home)

| References | Year | Contribution |
|---|---|---|
| Sourantha et. al. [139] | 2018 | Improved security of smart home automation system |
| Liu et. al. [140] | 2016 | Detects and prevents smart home billing system |
| Khan et al. [141] | 2016 | IoT Based approach for smart home energy utilization |
| Kang et al. [142] | 2017 | Framework for improving integrity and access control |
| Malche et al. [143] | 2017 | IoT enabled smart home system for controlled automation |
| Naik et al. [144] | 2018 | Open-source framework for communication reliability and security |
| Feng et al. [145] | 2017 | Intelligent and reliable smart home architecture for the improved living conditions. |
| Al-Kuwari et. al. [146] | 2018 | IoT based smart home controlling and automation |
| Singh et al. [147] | 2018 | IoT based smart home automation and security |
| Zhang et al. [148] | 2019 | Privacy oriented smart home architecture |
| Gajewski et al. [149] | 2019 | Possible cyberattack detection nd countermeasure for smart home |
| Ferraris et al. [150] | 2020 | Privacy-preserving smart home trust model |
| Guhr et al. [151] | 2020 | Smart home user health data security and privacy |

## 4. Smart City Examples Around the Globe

Barcelona, Amsterdam, New York, London, Paris, Singapore, Toronto, Japan, Hong Kong, Berlin, Reykjavik are cities that stand out according to Forbes Magazine [165], [166]. In 2012, being one of the first European cities, Barcelona started embracing IoT technologies to evolve into a smart city. Being the Mobile World Capital, the city mainly focuses on improving its transport sector. Barcelona first launched its energy-saving LED-based street lighting system in 2012. The sensors incorporated within the lighting system enable measurement of traffic density, weather, climate conditions, sound and air pollution levels in the environment, pedestrian activity, etc. The city has simultaneously worked on its waste management system, including vacuum-integrated waste bins that suck and accumulate trash underground to prevent waste odors and expand the waste storage capacity. This method helps decrease the visits of waste collecting vehicles, which reduces the noise and air pollution caused by these vehicles. Barcelona's intelligent parking system, also introduced in the same year, works by notifying and directing drivers to available parking spots, which reduces traffic congestions and emissions. Also, solar panels and hybrid buses in the bus transit systems help cut down emissions [167]. In 2016, the pedestrian-friendly Superblocks urban plan was first implemented in Barcelona in the Poblenou neighborhood. This low-cost urban concept aims to reduce air pollution and noise pollution caused by vehicles and waiting times for buses by restricting traffic and providing streets to be used entirely by pedestrians and cyclists. It also ensures that pedestrians do not have to rely on traffic lights and drastically reduces the risk of pedestrian-vehicle related accidents [168].

Recognized as the "Smart Nation," Singapore first declared its smart city initiative in November 2014 and is currently one of the topmost examples in the globe [165], [166]. Singapore has recently started the National Digital Identity, NDI initiative for its citizens and businesses, which can verify a user's identity online for secure online transactions. Apart from monitoring environmental factors, smart street lights sensors can also collect acoustic data such as screaming during an accident. The wide distribution of sensor networks all over the city allows the government to monitor almost everything, even people smoking in restricted zones or throwing trash out of their buildings. Nowadays, most Singapore residents have adopted e-payments or digital payments, which includes payment through QR codes and other mobile applications. After Singapore being the first city in the world to introduce autonomous vehicles, the Singapore government plans to launch self-driving buses, install energy-saving street lights on almost all public roads, and incorporate solar PV panels on top of the roof surfaces of around 6000 buildings all by the year 2022 [169].

Also being ranked among the top 10 smart cities according to IESCE Cities in Motion Index, and considered as the 3rd best city in terms of technology, Amsterdam listed first among European cities. [170]. As a part of the Amsterdam Smart City (ASC) circular city project, waste materials are reused to generate electricity and Carbon dioxide. Smart grids are used to convert carbon dioxide into energy. In transportation, though bicycles are widely used, there has been a rise in the use of Electric Vehicles, EV in Amsterdam city that allow energy exchange between the grids and vehicles. In the energy sector, a smart grid is used to generate energy from renewable energy sources, which can be stored or shared when required. Also, LED street lightings can conserve around 80% of energy while saving 130 billion euro [170].

Table 3: Top 5 Smart cities according to IESCE Cities Motion Index

| Smart City | Technologies Used |
|---|---|
| New York City [171] | • Automated Meter Reading (AMR) to track water consumption.<br>• Solar powered trash bins to measure trash levels<br>• Energy efficient smart lighting to reduce GHG emissions and costs.<br>• Web-based surveillance from HunchLab uses historical crime data to predict future crimes.<br>• Air quality monitoring |
| London [172] | • Street lamp posts equipped with air quality sensors, surveillance systems, electric vehicle charge points, public Wi-Fi<br>• Self-driving Heathrow pods<br>• London data store to make data accessible to all. |
| Paris [173] | • IoT enabled park benches.<br>• Halle Pajol solar power station – an industrial building equipped with solar panels.<br>• E-bike renting schemes. |
| Tokyo [174] | • Smart Grids to limit energy wastage<br>• Energy Management Systems in homes (HEMS), buildings (BEMS), and factories (FEMS)<br>• Tokyo Rinkai disaster prevention park that uses smart technologies to help citizens survive during natural disasters |
| Reykjavik [175] | • Public transportation app, Straetó for operating city buses.<br>• ON Power – Reykjavik Energy Company that uses geothermal energy to produce electricity.<br>• Reykjavik Fibre Network offers 100% Fibre to The Home, FTTH connectivity.<br>• The Better Reykjavik website allows citizens to share their opinions with the government. |

## 5. Security and Privacy Issues

These are the two crucial dimensions of any digital infrastructure. The smart city is a new addition to the sustainable digital society. Potential privacy and security concerns need to be discussed for its improvement. In this section of the paper, various paradigms of security and privacy issues, including the recommended solutions, are discussed [176].

## 5.1 Security Attributes of Smart City

The smart city makes life easier by increasing the privileges of people's daily life. Smart healthcare, smart education, proper security, intelligent transportation systems, etc. are the core smart city implementation applications. As people are connected in a distributed network, each person is dependent on each other. This kind of connectivity needs a robust system. The concept of a smart city puts forward a new era of human life. But it has also presented different types of crucial issues. The shortcomings in understanding the security and privacy measures will eventually be the reason behind the rise of hazardous events in the whole system. So, we have to know and understand the smart city's security challenges for implementing a proper and secure smart city [177,178].

### 5.1.1 Security in Communication

In a smart city, people are connected to a network. Smart cities' core intention is to share the data among the users, and the users collect the data and analyze the data for other privileges. So it is crucial to establish security. Encryption is a beneficial and widely used technique for securing the communication system. On the other hand, an IoT based crypto-manager device management system can be used for ensuring secure communication [179]. While a user gets into the network, IoT devices automatically detect the user device and authenticate it to ensure the secure communication process [180].

### 5.1.2 Bootable Device Security

A computer virus, threats, or other malware are self-executable programs that damage computer systems. Every system needs an operating system or a kernel to control the system. While installing an operating system via bootable devices, it is crucial to check whether the device is free from malware or pre-boot malware. Pre-boot malware is more dangerous as it is executable even if no operating system has been installed. Moreover, sometimes it is tough to detect pre-boot malware. For this reason, secure booting is one of the most critical requirements. Ultra-low-power hash function consumption is vital to implement for IoT devices [181].

## 5.1.3 Supervisory Security

The building block of smart cities is IoT devices. They are organized systematically to build a concrete environment that can provide necessary features to the citizens. If they are not appropriately monitored, cyber-attacks like fake sensor data or injection of erroneous data may harm the whole system. Moreover, analysis of malicious behavior and attacks must be performed properly to generate an automated response. In order to do this analysis and generate a response, the system must have certain scalability to let the IoT device process a large amount of data. A system can work in an elimination and response strategy to eradicate this type of suspicious behavior. The system must nullify, temporarily separate or withdraw the affected sections of the IoT device. Along with other monitoring strategies, organizations performing different operations in a smart city are also responsible for monitoring suspicious activity, the services' stability, tracking abnormal behavior, or any other system-threatening action with appropriate mechanisms.

### 5.1.4 Comprehensive City Lifecycle Security

The core of a smart city is the citizens. To collect data and information from citizens for implementing a complete system, IoT devices need to maintain a lifecycle enhancement. The lifecycle management of a system and application is a very complicated procedure as it needs to be done at different design stages. Moreover, the developers must investigate the code, critical issues, and other components related to the specific IoT devices before connecting them to the system. Smart City Comprehensive Data Life Cycle (SCC-DLC) is proposed to manage the enormous data size [182]. This kind of approach also reduces communication latencies and traffic preventing system failure. Considering the process of data collection from various resources, storage, and data acquisition, it works fine to maintain and enhance the lifecycle of a system and application in a smart city [183].

### 5.1.5 Updates and Patches

Issues can arise as time passes. So regular updates are essential to defeating vulnerable attacks on security. Authenticating the update patches is a significant addition for IoT devices. These established patches are generated and transmitted by their service providers and operators. Companies have to develop patch compliance policies, proper testing, and no performance compromise for the devices and bandwidth efficiency [184]. There might be some challenges for updating the patches also. For instance, medical applications devices cannot endure antivirus as a security solution [185].

### 5.1.6 Access Management and Information Security

IoT devices in a smart city work in a way where they need to generate and transmit data continually. At the same time, this data is managed and controlled; the system must prevent unauthorized access. The system is a smart city that must be powerful enough to identify users, authenticate activity, and control data use. Legal citizens' access to data building secure communication is the essential precondition for the authentication process. It will ensure a smart city environment to resist downtime, unexpected changes, and tampering. For a concrete communication system and ensuring data privacy, access control, and authentication protocol like

Role-Based Access Control (RBAC) [186], Identity Based Encryption (IBE) [187], and Attribute-Based Encryption (ABE) [188] are proposed.

## 5.2 Leveraging Security Issues

Smart cities bring convenience, flexibility, efficiency, and automation to our life. The implementation is still conceptual. The concept of a smart city is to make everything interconnected through a connecting medium. This connectivity surely needs a state of the art infrastructural design to better security and privacy solutions. As smart cities are heterogeneous devices, security threats are also diverse. For this reason, all types of attack vectors can be considered as potential security threats for a smart city. Confidentiality, authenticity, availability, and integrity are major concerns of smart city data security. Confidential information is very much crucial for user data. This protects the data from being leaked to unwanted parties. The integrity of data means the correct and non-modified data. If the data is modified or damaged while transmitting and storing, it might produce the wrong results. This false result will produce incorrect decisions [189]. This section details some specific security and privacy issues for smart cities and potential solutions where available.

### 5.2.1 Cyber Attacks

Data needs to be available always for all the data subjects. The availability of data is a major concern in the case of data processing and analysis. There can be various cyber security attacks on data. Besides, different type's attacks can be used to (i) gain access to people's confidential information, (ii) monitor user activities (iii) make the system unavailable to users.

A smart city comprises various interconnected IoT devices. As the smart city application vector increases, the amount of interconnected IoT devices also increases. Someone with malicious intent could breach any of the devices and begin spreading malware across the entire network. This malware can be used to monitor the whole network, eavesdrop, and sniff important data from the network. Due to this attack, confidentiality, authenticity, and availability of the data can be compromised. For this reason, malware detection and mitigation techniques can be implemented as a solution to this type of threat [190].

An example of another cybersecurity attack is the Distributed Denial-of-Service (DDoS) attack. This is increasingly common nowadays. DDoS attacks could be used to bring down an essential part of the smart city system, for example, traffic lights and cameras in a smart transportation system. This type of attack will compromise the availability of devices and data in smart cities. Attack detection and mitigation tools can be used as a countermeasure of this attack. While designing a countermeasure, the nature and attribute of a smart city must be considered so that the solution can be implemented in a growing network and various types of devices. A possible solution is proposed that includes a novel framework using Software Defined Networking (SDN). It can detect a DDoS attack and differentiate it from a sudden increase in the number of legitimate users [191].

The residents of smart cities use various types of web applications and services. Authentication data, as well as personal data, is used for registering and using the services. For example, social networking sites are handy for maintaining social connections and coherence among the community. This site stores user data, and analyzing social networking data can sometimes reveal critical human behavior. Cross-site scripting attacks can exploit web applications and services. This type of attack compromises the authenticity and confidentiality of the data. This type of attack can sun malicious scripts on web pages. SQL injection is another harmful attack that can exploit the database vulnerability to read useful authentication information. The attacker can modify any other data residing in a database table. There are widely accepted and popular countermeasures against this type of web application attacks [192]. Blackbox testing [193], static code testing, input type checking are some of the techniques to countermeasures of this type of attack [194] [195].

### 5.2.2 Devices and Connectivity Threat

The smart city comprises smart devices, e.g., smartphones, smart healthcare devices, etc. These devices have the relatively smaller processing power and storage. For this reason, data processing at this end is very minimal. These smart IoT devices use third-party applications and services for their operation, which often is the point of vulnerability. Besides, the nature of the devices poses various security threats irrespective of their current state of operation. For this reason, effective countermeasures have been proposed for these attacks [196] [197]. It would be really difficult for an IoT-based smart city to establish a cryptography method for secure end-to-end communication. One of the reasons for this is that the devices are all made by different manufacturers and thus held to various security standards. Furthermore, a lot of the devices will have very low computational power. To combat this issue, IEEE 802.15.4 has been put forth [198]. However, this protocol has some security flaws. It is a daunting task to provide proper security measures considering the smart devices' nature and attribute [199].

### 5.2.3 Security Issues in Data Collection and Transmission

Smart cities are largely dependent on data that is collected from various sensors and other IoT devices. This vast amount of data needs an efficient and secure storage facility. To accommodate this increasingly large amount of data, and auto-tiering database facility is proposed to be used. This technology dynamically allocates data storing facilities according to the organization's rules and regulations set by the administrator. Besides, dynamic storage location assignment poses vulnerabilities. For example, proper cybersecurity techniques are not integrated with it. On the other hand, data can face collusion threats. So that more than assigned or the storage providers can access permitted data. Data storage and transaction creates a database history. This history reveals useful information that can be harmful if it goes to the wrong hand. So, data protection and security mechanisms for this log data is also a crucial need [200].

Smart cities aim to provide state-of-the-art real-time facilities to the citizens. To facilitate this purpose, real-time data analysis is needed. As the data source is heterogeneous and huge, it is challenging to transmit new, authentic, complete real-time data. Sometimes due to spoofing attacks, the data can collude. The attacker can create multiple fake identities so that fake and unreliable data can be transferred to the processing unit. If such data is provided to the critical organs of a smart city that operates on real-time data, it will create a massive problem for the citizens as well as for the city infrastructure. For this reason, efficient attack mechanisms against this type of attack (sybil attack, spoofing, etc.) can be used for the smart city devices that collect data.

### 5.2.4 Physical Attack on the Infrastructure

Apart from data protection in the virtual world, physical infrastructure security is also crucial for security purposes. Wired and wireless sensors are used in infrastructures. Smart homes, smart buildings, smart transportation infrastructure, etc. use sensors and other versatile IoT devices. Sensors devices can be damaged easily if intruders can access physically. Moreover, in smart homes, smart buildings, IoT devices can be fed with fake data to provide wrong results [201]. Wired transmission lines can be tapped to sniff transmitted data. Intruders can inject malicious code into the sensors and reprogrammed it to operate and provide falsified data. For this reason, physical infrastructure security is a must for smart city security solution purposes. This will help to authenticate the verified source of data and data integrity in terms of unaffected devices [202] [203].

### 5.2.5 Cloud Related Security Issues

Cloud is introduced to the smart city due to the resource restriction for application hosting, storage, processing, etc. These issues are already associated with IoT enabled systems. Since IoT is the basis of the smart city concept, it needs to be addressed. For leveraging the issue, the cloud has come into effect. Cloud computing [204] is a shared computing service in a virtual environment. It facilitates a variety of services to persons and enterprises. As cloud computing supports multi-tenancy, multiple users can use cloud services for different purposes. This raises potential issues of malware and malicious code injection in the system. This will hamper the shared data security. Malicious code injection can cause identity theft, unauthorized device or storage access, data sniffing using man in the middle attack. As users' data and services are at risk, adequate security measures against this type of cloud security vulnerabilities need to be implemented [205] [206]. Several pieces of literature discuss various aspects of security solutions. Since smart cities comprise various technologies, their security solutions are also different methods and protocols. Some of the literature that discusses security solutions are given below in a tabular form.

Apart from the studies mentioned above, there are some other notable recent researches in smart city security. Smart services, including power supply, proper water supply, communication services, etc. are crucial for a smart city to run smoothly. Unauthorized access inside these services can potentially hamper the services and hamper the citizens' quality of life. Toh et al. [207] describe the importance of these services and how they can be vulnerable to attackers. Potential countermeasures mechanisms, including the use of firewalls, cryptography is also elaborated. Xu et al. [208] talked about the DDoS attack strategy on smart cities' data management services. DDoS attacks can exploit the availability of services and devices. Hence, a defense mechanism was proposed, which uses network traffic monitoring and classifying mechanisms. Blockchain is a solution factor for various security vulnerabilities.

Similarly, smart city issues and challenges are also addressed using blockchain technology. Hakak et al. [209] proposed a blockchain-based secure architecture for smart cities, which provides an in-depth understanding of various smart city case studies. A secure transportation system is vital for smart cities. A distributed system is proposed for intelligent road transportation (toll pricing) by Bouchelaghem et al. [210]. This study simulates various attack scenarios, countermeasures performance analysis of the proposed system. Smart grid powers up the city; hence the importance is infinitesimal. Islam et al. [211] proposed a secure framework for smart grid security. This framework secures the physical layer of the smart grid using artificial intelligence. Similarly, another physical layer security approach was mentioned by Wang et al. [212]. Mohammad [213] proposed a holistic approach to secure a smart city. In his paper, critical smart city infrastructures were identified to determine the efficient approach for securing smart cities. Castillo-Cara et al. proposed an Application-based approach for protecting women against violence [214]. This is a citizen-oriented public safety application that uses constant location monitoring for securing smart city citizens. Akhunzada et al. [215] depicted a detailed explanation of IoT enabled device usage, importance, vulnerabilities, and countermeasures. This taxonomical approach addressed 5G cellular technology for smart city service improvement.  Another holistic overview of crucial security issues and vulnerabilities of different smart city infrastructures and services are described by Ismagilova et al. [216]. This study mentions various security frameworks of smart city services and infrastructures. It also briefly describes a blockchain-based approach for securing smart cities.

### 5.3 Privacy Issues and Solutions

Privacy breach [217] and protection is a long-standing topic in every aspect of technology. As smart city residents are continuously connected with the network infrastructure, it is imperative that the location can be tracked at every moment. Location information can reveal a person's residents, office, frequent place of visit, friends place, etc. [218]. Social networking sites have various features, including posting pictures, sharing locations, videos, etc. These can reveal personality traits, next movement prediction, behavioral anomaly, etc. So, a person's social data is another essential factor of privacy issues [219] [220]. Smart cities use surveillance cameras and various sensors to detect the activity of a person. This helps in the smart home, office, and industry activities for energy consumption purposes too. Moreover, every day, people upload videos of their moments in photos, videos, animations, etc. These contents spread across the network in a minute. For this reason, privacy should be addressed in terms of monitoring activities, media sharing, and consent management techniques, too [221].  Smart city privacy issues can be categorized according to the following -

### 5.3.1 Public Data Access Management

To provide better services fulfilling citizen's requirements, data plays a vital role. Organizations like the government need to access this data for different needs, like ensuring national security. But open usage and access to other applications of third parties may lead to identifying users. The amount of data that is stored for smart city applications is enormous. So, it is challenging to handle all the data at once. Data may also be sometimes misleading due to randomization or generalization. Applications may conceal an individual's data to use for their purpose. This also reduces privacy. To meet privacy concern, raw data should be hidden from users. This can be done using cryptographic protocols like homomorphic encryption [222]. It allows the user to maintain anonymity from the application. There are some practical examples of this issue. Almere uses the StraatKubus platform that allows only employees of the municipality to access. Cities like Sydney, Chicago, and Hague also follow some approaches to mitigate this privacy concern [223].

### 5.3.2 Widespread Communication

Smart cities are built upon the infrastructure of communication among various devices. It can cause issues; for example, an individual's location can be discovered through a public hotspot. Besides, unencrypted traffic from these unprotected public hotspots can be read by other parties by gaining access to wireless channels. Only secure connections cannot ensure data safety. Hackers can use metadata leaked from this public hotspot to reveal certain information from them. Every mobile device has a large number of sensors that can be used to monitor the user's activity. Again, these sensors and other small mobile components are developed and controlled by different groups [224]. Moreover, most of the applications in mobile devices are produced by third-parties. These applications often want to access more information than needed.

The best way to prevent privacy leakage is to ensure the wireless system is encrypted. WPA2 is such an encrypted wireless system. Moreover, there are SSL, or TLS approaches to secure mobile apps. There are also approaches like anonymous communication using Onion Routing like Tor [225], which are very important in health services sectors. But these tools are very problematic for users with no previous learning. Software that is already installed and configured can solve this problem. Still, there are possibilities for attacks like traffic correlation and timing attacks. Fingerprints, iris, retina, pulse rate, etc. can be used for maintaining device privacy. Changing device identifiers frequently, randomizing, and inserting cover traffic can help protect websites and mobile devices [226].

### 5.3.3 Monitoring Devices

Personal data that are generated monitoring devices can be minimized using isolated sensors. Moreover, raw data can be ignored. Sensors can also use aggregated data like histogram, count, etc. to imply the application's purpose. K-anonymity can ensure location privacy using some attributes like location and Time of reading, Time, and area changes until k individual readings. An anonymous server can serve the purpose, but it must be trustable. Sensors will only look into location when the user permits to use the information. Otherwise, it is not possible to hide the location from public uses. In some, it is challenging to use k-anonymity for the devices not always able to connect. The tessellation approach can be followed to overcome this where the area's size is pre-computed [227]. K-anonymity is not trustable for location information. In those cases, l-diversity is followed where a certain area remains hidden containing at least l points of interest. Furthermore, collusion between data nodes can be restricted by splitting them into different nodes [228]. Cities like Glasgow, Rio de Janeiro, Eindhoven, etc. have developed smart cities' features.

Devices like CCTV, intelligent light, traffic light, parking facilities are maintained facing this kind of issue. Data aggregation to the municipality, third parties, are some approaches to mitigate the issues [229].

### 5.3.4 Wearable Devices

There are multiple wearable devices like a smartwatch, smart glass, etc. embedded in a smart city network. They also face some privacy threats. The apps or programs that run into these devices may sometimes have flaws in their architecture and functions. They may set to reveal critical personal data. The communication channel and third-party servers are often vulnerable to hackers and disclose sensitive information. It is also prone to attacks like routing or man-in-the-middle attacks [230]. Devices need to be permitted for offline operation. It needs to have the mechanism of local data processing. Moreover, storing a minimum amount of data can decrease privacy issues. Such as it can store only the minimum duration rather than the whole timestamp of an event. This will reduce the use of storage and increase efficient operation [231]. Again, homomorphic encryption is a must for a service provider. It will restrict providers from reading personal data while processing. The main reason for this type of issue is resources. Firstly, sensors of the network need to be secured to prevent intrusion. The encryption key for the data encrypted must be stored in secured distributed storage that follows role-based access control [232] [233].

### 5.3.5 Use of IoT

IoT has provided many smart city features like smart homes, smart buildings, smart transportation, and many more. It has to operate on a large amount of data. So, it needs a strong security concept. Moreover, the system must ensure that one agreement does not affect the whole system. Hence, some issues need to be mitigated for strong privacy. As there is an enormous size of data to be handled by IoT devices, the transmission of processing is very crucial. The communication between IoT devices can be affected by several data attacks. Without proper security measurements and management, malicious data can be injected. This can eventually destroy the whole system. There is also the possibility of leaking sensitive data during operations. For every service provided by a smart city, IoT devices need to operate on the owner's name and type of service. These essential data can be passed to intruders during service discovery revealing the physical identity of the user.

Service like smart grids needs to operate on a smaller part of data for efficient operation and reduce utility loss. Trusted third-party, homographic encryption, and bilinear maps can be some promising data totalization approaches on this remark. Again, cryptography can enable local operation correctly. This lets the user know more accurately, for which purpose their data is being transmitted and used. It is a must to ensure that valid clients can only reveal a user's identity only when the user wants to prevent data leakage. Private Service discovery lets the users have the opportunity to reveal their identity only when they are connected to the correct service. The service policy is encrypted with identity-based encryption ensuring mutual authentication [234 -237].

### 5.3.6 Smart Card

A smart card is a technology that is used for several features of a smart city. For example, public transport, tracing, personalizing, advertising, etc. have extensive use of smart cards. The main issue of the smart card is the logging of transactions. The travel history can reveal many things regarding locations, visits, food habits, etc. These data are meaningless alone. But linking many data, a particular relation can easily be defined.

Splitting service and the transaction can be a possible solution. When user authentication is separated from service access, it will be hard to imply any relationship between user behavior and transaction history. This process can be implemented by an anonymous smart card that does reveal any type of identifying information. Following attribute-based encryption, the user's identity can be verified without revealing it [238]. There are also ways to disclose a user's identity even if the system can unlink services from a transaction. Only information about origin and destination is enough to reveal an individual's identity. To solve this, we need to reduce the amount of data that the system will process. There should be a policy for which data will be stored and not [239] [240].

### 5.3.7 Transportation Technology

Another essential feature of the smart city is intelligent vehicles. Along with this, several transportations and traffic systems are also introduced for better features. But they also have privacy issues. Vehicles need to share information like speed, direction, position, turn signals, identity, etc. to avoid accidents. Other vehicles make their move based on this instruction, and a rule is imposed automatically in the traffic system. But this information transmitted to everyone is unencrypted. Anyone inside the range can access the information and learn all the status of other vehicles. This approach is revealing the identity openly. The metadata that any vehicle poses can be used to reveal its true identity. It also lets the attacker trace if it is in the transmission range.

This malicious transport should be removed to ensure overall security. This is done by backward-privacy that refers to the cancellation of all the identifiers of that specific vehicle. It removes short-term identities for the future of that vehicle. Rather than transmitting the real identity, vehicles use a pseudonym. Moreover, to maximize privacy protection, all the vehicles need to change short-term identifiers together [241]. Vehicles need to get in touch with many other services. The data generated from these services are enormous. The cryptographic approach is a must to ensure these data's safety [242] [243].

### 5.3.8 Service Provider

A smart city is all about data and their feedback creating multiple features for the citizens. The service provider handles this data. They store, process, and work upon this data. The data service provider can see the data stored in the cloud. So that all the stored data must be encrypted. Service providers can't decrypt them without authentication. Moreover, there are encryption processes like attribute-based encryption, where data can be stored with multiple attributes [244]. The process, like attribute-based encryption, provides users many facilities. Users can authenticate data using these attributes. Also, cloud providers cannot keep track of the activity of users [245].

Some recent studies have shown crucial privacy issues and solutions. Santana et al. [246] the rapid growth of the population in smart cities. Increased population means increased communication and monitoring. The author proposed a privacy-preserving monitoring system of crowds in various infrastructures using artificial intelligence and wifi-based tracking. This system provides an efficient crowd management strategy using wifi data. Xiao et al. [247] discussed data privacy in the case of seamless data transmission among companies. The author here proposed an automated privacy management system for data exchange in case of cross-industrial data exchange. Software updates are widespread for every system. Generally, a system needs to shut down entirely to implement new updates which are not quality of service-oriented. Moreover, data privacy is not preserved for outdated software.

For this reason, a sophisticated system that updates software without shutting down the whole system is proposed by Mugarza et al. [248] In addition, data privacy is addressed using homomorphic encryption technology. UAVs are important for surveillance purposes for ensuring security. It collects a lot of personal information as it moves across the city. For protecting the privacy of the citizen's data, Kim et al. [249] proposed a privacy-preserving strategy. Using this proposed strategy, the device will make minimal movements to collect less personal data. Zhang et al. [250] designed a recommender system that addresses user data privacy. This system uses an anonymous address to hide the reviewer's original identity and provide a rating matrix for this commander filtering strategy. As IoT is an integral component of smart city architecture, Badii et al. [251] depicted a brief description of privacy issues and solutions in IoT for a smart city. A Similar Approach is followed but illustrated a border sense by Al-Turjman et al. [252]. The authors described smart city applications and state of the art privacy and security issues and solutions for IoT based smart cities. Braun et al. [253] and Habibzadeh et al. [254] elaborately discussed a holistic overview of privacy and security issues and challenges in a smart city, their solution, and challenges. Habibzadeh et al. additionally described policy-based issues and challenges of a smart city. Like other issues, challenges, and applications, a blockchain-based architecture has been proposed by Sharma et al. [255] that addresses network architecture and privacy issues and solutions of the smart city.

## 6. Future Research Directions

The concept of a smart city is still emerging. The development and progress is going on in various parts of the world. Though the developed countries were the ones to materialize the smart city idea, many developing countries are on the same floor also [256]. The existing works bring hope to this smart city concept though extensive research is needed to spread adoption and application. Some of the notable future research direction can be as follows-

### 6.1 Blockchain for Smart City

To solve privacy concerns, most of the smart cities have considered using Blockchain technology as it is pseudonymous. Several schemes regarding anonymous Blockchain-based systems have also come forward. However, they are yet to have a successful model and address certain other challenges. The privacy aspect of smart cities is definitely up for further research. Blockchain brings immutability, anonymity, and confidentiality. Some of the data security issues can be solved using blockchain [257]. On the other hand, blockchain and smart contract enabled applications can also leverage automation issues. The transaction is trusted so that transparency will increase among the stakeholders [258, 259]. A considerable number of data is continually being generated. Protecting the privacy of this large number of data requires a robust cryptographic algorithm. Working on suitable cryptography mechanisms and algorithms needs more attention as well.

Besides, blockchain can be used for big data to address big data security, confidentiality, and integrity in the future.

## 6.2 Cost management

This is crucial for any sector; a smart city is no different. Increased operational cost and design cost can strengthen the development of a smart city. For development purposes, a city needs infrastructural development as well as improve the design perspective. On the other hand, budget optimization is crucial for maintaining the economic stability of the community. This area is a growing research concern for the organization considering the complexity of running the smart city's diverse sectors and providing QoL, and protecting sustainability. Most of the smart city domains have been proposed to implement with the help of Blockchain. Some deep learning techniques can be sought after for predicting future costs required for a smart city. Targeted experiments can be carried out on the existing smart city test-beds.

## 6.3 Sustainable Energy Management

Energy is a growing concern for future cities. Renewable energy sources can be one of the most innovative and suitable solutions for protecting sustainability. This approach shall reduce environmental pollution and reduce the financial burden for energy management purposes [260]. Some initiatives have been taken for energy efficiency in the smart grid with the help of Blockchain. Blockchain has many consensus mechanisms for handling this aspect. However, they lack scalability. A new consensus algorithm, Proof of Trust, claims to address the scalability issue but is still in its infancy. [261] Other technologies like IoT, CoT, edge computing, etc. can be integrated with Blockchain to provide scalable and energy-efficient platforms. Research can also be done on the construction of smart energy that will be economically sustainable as well.

## 6.4 Data Storage and Processing

A smart city comprises various devices that accumulate a huge amount of data. Data needs storage and processing to make it useful decision making or prediction. Therefore, integrating Big Data analytics into smart cities has become a crucial goal. Several studies have been conducted on incorporating a scalable data storage system but are mostly still in the research phase. They were mostly done using Blockchain, but the chain can become quite bulky, so it's impossible to integrate Blockchain into a system directly. Several works have been proposed using an off-chain data storing system [262] and BigChainDB [263]. Other technologies like Machine Learning, Tactile

Internet, Cloud Computing, etc. can also be used in smart cities to address data storage issues effectively. However, Cloud Computing and fog computing can be a costly approach [264].

## 6.5 Emergency Services

Constructing a smart emergency response or smart ambulance is yet to be done and could be a possible future direction. Telemedicine platforms could also use technology like Blockchain, IoT, and other telecommunication infrastructures to improve their service. Smart transportation also has scope for improvement. Traffic signal management or traffic congestion management could use more research to create perfect smart traffic solutions. Most of the smart transportation work involved roadways. Airways or waterways can be considered as research areas. The smart ticket system is also an interesting topic and beneficial for the whole smart transportation industry. Image or video processing techniques, artificial intelligence, and Blockchain can be used to deploy a complete smart transport system that can use real-time analysis. Another growing research field in terms of smart cities is smart education. Much work has been done on this topic, but further studies can be helpful. Use of augmented reality, virtual reality as well as IoT, Blockchain can create a better learning environment and further improve the outlook of smart cities. A lot of such ideas are still on the idea-level. With enough research and experimentations, deploying a full-functional and smart real-world town [265].

## 7. Conclusion

The smart city combines the core concepts of sustainability and smart living. Though the implementation of a complete smart city environment is yet to be seen in any part of the world, several smart cities are already being used by various urban regions. The idea can promote not only an admirable living quality but also ensure public safety and security. Researchers are trying to put forth their innovative ideas for making the concept bulletproof before it comes into play as a complete package. State of the art facilities and application integration with different services increase security and privacy risks. Thus, a proper smart city physical infrastructure that can meet the citizens' demands and defend the citizens from any unexpected incidents is needed.